\documentclass[aps, pra, twocolumn, superscriptaddress]{revtex4-1}
\usepackage[pass]{geometry}
\usepackage{graphicx}
\usepackage{amssymb}
\usepackage{amsmath}

\usepackage[usenames,dvipsnames]{xcolor}

\newcommand{\subscript}[1]{\ensuremath{_{\textrm{#1}}}}

\begin{document}


\title{Inversion of Zeeman splitting of exciton states in InGaAs quantum wells}


\author{P.~S.~Grigoryev}
\affiliation{Spin Optics laboratory, St. Petersburg State University, Ulyanovskaya 1, 198504 St. Petersburg, Russia}

\author{O.~A.~Yugov}
\author{S.~A.~Eliseev}
\author{Yu.~P.~Efimov}
\author{V.~A.~Lovtcius}
\author{V.~V.~Petrov}
\affiliation{Department of Physics, St. Petersburg State University, Ulyanovskaya 1, 198504 St. Petersburg, Russia}

\author{V.~F.~Sapega}
\affiliation{Ioffe Physical-Technical Institute, Russian Academy of Sciences, 26, Politechnicheskaya, 194021, St-Petersburg, Russia}

\author{I.~V.~Ignatiev}
\affiliation{Spin Optics laboratory, St. Petersburg State University, Ulyanovskaya 1, 198504 St. Petersburg, Russia}

\date{\today}

\begin{abstract}
Zeeman splitting of the quantum confined states of excitons in the InGaAs quantum wells (QWs) is experimentally found to strongly depend on the quantization energy. Moreover, it changes its sign when the quantization energy increases with the decrease of the QW width. In the 87-nm QW, the sign change is observed for the excited quantum confined states, which are above the ground state only by a few meV. A two-step approach for the numerical solution of two-particle Schr\"{o}dinger equation with taking into account for the Coulomb interaction and the valence-band coupling is used for theoretical justification of the observed phenomenon. The calculated variation of the $g$-factor convincingly follows the dependencies obtained in the experiments. 
\end{abstract}


\maketitle

\section{Introduction}

Magnetic properties of Wannier-Mott excitons are extensively studied since their first observation at the adsorption edge of Cu$_2$O crystal~\cite{Gross, Zakharchenya}. Recent enhanced experiments and a theoretical analysis have discovered a rich energy structure of the excitons in this crystal \cite{nature2014, Kavokin-Nature2014, Glazov-PRL2015}. Fine exciton structure is mainly determined by spin properties of carriers forming the exciton states. During the last two decades, the spin properties of exciton complexes have attracted considerable attention due to overall interest in spin physics and their possible applications in the field of spintronics~\cite{Awschalom-book2002, Dyakonov-book2008, Suter-book2013}. A major part of experimental studies has been fulfilled for the III-V heterostructures grown by the molecular beam epitaxy. In particular, the InGaAs/GaAs and GaAs/AlGaAs quantum wells~(QWs) were studied due to their high quality allowing experimental observation of the fine energy structure of excitons ~\cite{Iimura-PRB1990, Potemski-PRB1991, Snelling-PRB1992, Traynor-PRB1995, Traynor-PRB1997, Kotlyar-PRB2001, Chen-APL2006, Kubisa-PRB2012, Seisyan, Arora-JAP2013, Jadczak-APL2014} (earlier works are reviewed in Ref.~\cite{Bauer-PRB1988}). In these works, the magnetic field is applied along the growth axis and Zeeman splitting of the one or several lowest states of excitons are studied using various experimental methods. The splitting, $\Delta E$, is discussed in terms of exciton $g$-factor, $g_{ex}$, defined by relation: $\Delta E = g_{ex}\mu_B B$ where $\mu_B$ is the Bohr magneton, $B$ is the magnetic field, and $\Delta E$ is the distance between exciton states active in $\sigma^+$ and $\sigma^-$ circular polarizations.

It has been found that the exciton $g$-factor strongly depends on the QW width~\cite{Snelling-PRB1992, Traynor-PRB1995, Traynor-PRB1997, Kotlyar-PRB2001, Chen-APL2006, Arora-JAP2013, Jadczak-APL2014} and on the magnetic field magnitude when it exceeds several Teslas~\cite{Potemski-PRB1991, Kubisa-PRB2012, Jadczak-APL2014, Yasui-PRB1995, Timofeev-JETPLett1996}. In particular, the inversion of the exciton $g$-factor measured in small magnetic fields has been reported in Refs.~\cite{Snelling-PRB1992, Kotlyar-PRB2001, Chen-APL2006, Arora-JAP2013} when the QW width varied from a few nm to a few tens of nm. In Ref.~\cite{Chen-APL2006}, the $g$-factors in a 20-nm Al$_{0.02}$Ga$_{0.98}$As/AlAs multi QW structure are reported to vary in the range from $g_{ex} = 0.5$ to $g_{ex} = -11$ for different excited exciton states. This variation is non-monotonic in energy of the quantum-confined exciton states.  
The exciton $g$-factor variations have been attributed to the variation of the hole $g$-factor because the electron $g$-factor weakly depends on the QW width~\cite{Yugova-PRB2007}. 

The physical origin of the hole $g$-factor variation is supposed to be the coupling of the heavy-hole and light-hole states~\cite{Iimura-PRB1990, Snelling-PRB1992, Chen-APL2006, Kubisa-PRB2012, Bauer-PRB1988, Durnev-PhysE2012, Durnev-FTT2014, Semina-Semicond2015}. An admixture of the light-hole exciton states obeying a huge $g$-factor~\cite{Durnev-PhysE2012} may considerably change the heavy-hole exciton $g$-factor. 

Large variation of exciton $g$-factor for different quantum-confined exciton states has been experimentally observed in several heterostructures with wide QWs~\cite{Kochereshko-PRL2006, kochereshko.1, kochereshko.2, kochereshko.3}. Effect of the quantum confinement of excitons in the QWs gives rise to quasiperiodic peculiarities in optical spectra  corresponding to the quantization of the center-of-mass exciton motion~\cite{Tredicucci, Ubyivovk}.  For such QWs, interfaces do not considerably affect magnetic properties of excitons, which remain similar to those for bulk crystal. This fact strongly simplifies theoretical analysis. The $g$-factor modification has been treated as the mixing of the relative motion of electron and hole in the exciton and the motion of the exciton as the whole~\cite{kochereshko.1, kochereshko.2}.

The quantum confinement stronger affects the exciton states when the QW becomes narrower. Theoretical analysis of excitons in such QWs should consider an interplay of the square QW potential and the Coulomb potential. Such consideration is rather simple for the case of a relatively thin QWs, which width does not exceed the exciton Bohr radius ($L < 15\,$nm for the GaAs-based heterostructures). For these QWs, the Coulomb potential can be treated as a perturbation, compared to the quantum confinement effect~\cite{Snelling-PRB1992, Chen-APL2006, Bauer-PRB1988, KiselevMoiseev, Durnev-PhysE2012, Timofeev-JETPLett1996, Ivchenko-book}. The problem of exciton magnetic properties in QWs of intermediate width ($15 < L < 150\,$nm for the GaAs), which are suitable for many applications, is more complicate. There is no analytical solution for such QWs~\cite{2015arXiv150800480K}.

In this paper, we experimentally study Zeeman splittings of several quantum-confined exciton states in the intermediate-width InGaAs QWs. We also provide a theory describing the Zeeman splitting of the ground and excited exciton states in the QWs. The theory is based on the numerical solution of the Shr\"{o}dinger equation for an exciton. The numerical approach is performed in two steps. First, we obtain a separate system of wave functions for the heavy-hole and light-hole exciton states and then we take into account the hh-lh coupling. The coupling of the unperturbed states is accounted according to the Luttinger Hamiltonian for degenerate valence band of the GaAs crystal. The comparison of the calculated Zeeman splittings with those found from the experimental study of exciton excited states shows convincing agreement of the obtained results. In such a way, we verify that large change of exciton $g$-factor with number of the exciton quantization level in the intermediate QWs is really caused by the coupling of the heavy-hole and the light-hole exciton states. The good agreement with the experiment allow us to understand, which interactions are mainly contributing to the $g$-factor modifications.

The rest of the paper is organized as follows. In section~\ref{experiment}, we present results of experimental study of exciton photoluminescence (PL) in magnetic field. These experiments are then discussed in terms of a theoretical approach described in section~\ref{theory}. In section~\ref{modeling}, important details of numerical solution of the the Schr\"{o}dinger equation are given. Then we discuss a universal character of the $g$-factor renormalization effect for QWs with different thicknesses. Conclusion section sums up the main results of our study.

\section{Experiment} 
\label{experiment}

\subsection{Photoluminescence at zero magnetic field}

We studied GaAs/InGaAs nanostructures grown by molecular beam epitaxy (MBE) technique. Three samples containing InGaAs layers surrounded by GaAs barriers have been grown. First sample, P554, contains the QW with nominal width 95\,nm and indium concentration of about 2\%. The second sample, P592, contains three spatially separated QWs with the nominal widths 30, 36, and 41\,nm and the indium concentrations 4, 5, and 7\%, respectively. Finally, the third sample, P531, contains four separated QWs (4, 7, 10, and 12\,nm with 5\% of In). The layer thickness in all the samples has a gradient, therefore the actual width of the wide QWs (wider than 30\,nm) was determined from the microscopic modeling of the exciton spectra (see details in Ref.~\cite{arXiv-asymmetric-QW}). For the points on the samples where the magnetic measurements were made, the fitted values of width are 87, 33, 40, and 45\,nm for the 95-, 30-, 36-, and 41-nm QWs, respectively. The samples were cooled down to the liquid helium temperature. 

To obtain the exciton energy positions, we have measured the PL spectra of the sample rather than the reflectance spectra because of simplicity of the experimental technique and the analysis of the spectra. Similar technique has been used in several publications, see, e.g., Refs.~\cite{kochereshko.1, kochereshko.2, Tredicucci, Tuffigo, Kusano, Arthur}. The PL was excited non-resonantly using a Ti-sapphire or a He-Ne laser. Spectral resolution of the setup was sufficiently better than the typical width of features in the PL spectra. We discuss the data for the first sample only in this section since the results are similar for all studied samples.

\begin{figure}[]
\begin{center}
\includegraphics[scale=1]{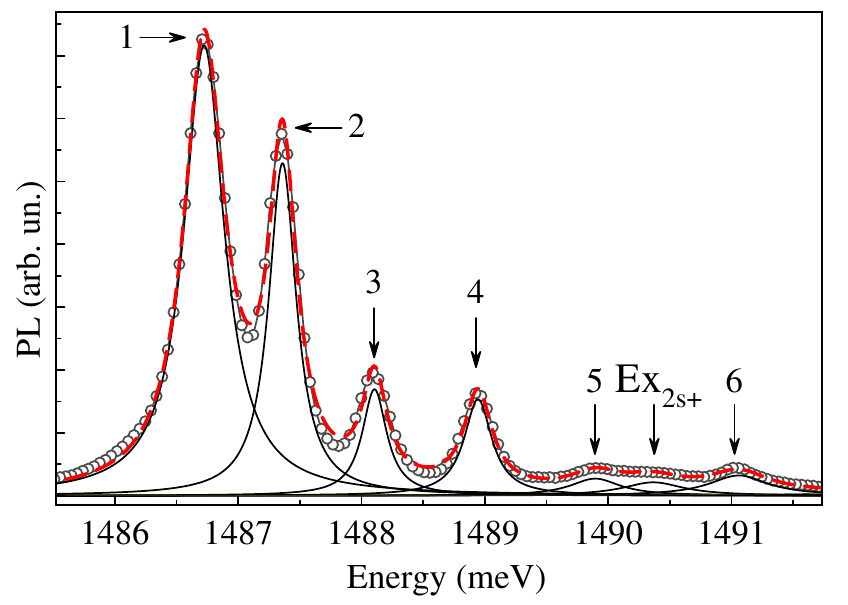}
\end{center}
\caption{Typical PL spectrum of the 87-nm QW in sample P554  (blank circles). Data fits are shown by the thin black Lorentzian contours. Exciton states are numbered; Ex\subscript{2s+} is the PL originated from the 2s and higher hydrogen-like exciton states.
\label{Flo:figure1}
}
\end{figure}

The PL spectra were fitted with a series of Lorentzian contours, as shown in figure~\ref{Flo:figure1}. The small width of peaks demonstrates high quality of the sample and leaves no doubt in the spectrum interpretation. The physical origin of the peaks is the PL of the quantum confined exciton states~\cite{Arthur}. Due to the high radiative rate and small energy distance between the states, the exciton thermalization is suppressed and the hot PL is observed.  The feature marked by Ex\subscript{2s+} is supposed to be the PL of the excited s-like exciton states.

\subsection{Photoluminescence in magnetic field}

Magnetic field effects were studied in Faraday configuration, i.e., the magnetic field was parallel to the excitation axis and perpendicular to the QW plane. The measurements were done with separate detection of PL in the $\sigma^+$ and $\sigma^-$ polarizations. 

\begin{figure}
\includegraphics[scale=1]{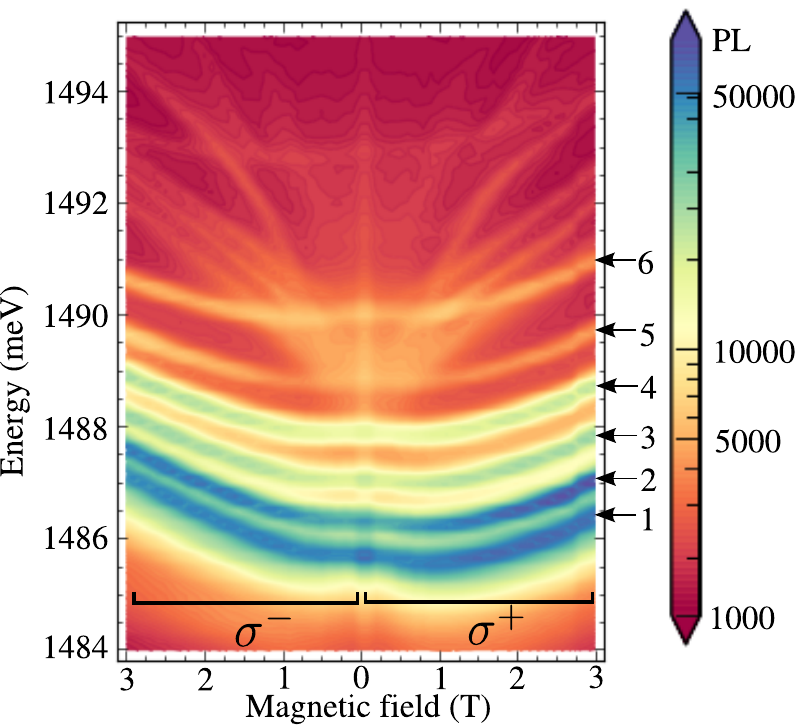}
\caption{PL spectra of sample P554 as a function of magnetic field for the left-handed ($\sigma^-$) and right-handed ($\sigma^+$) circular polarizations at $T=5~K$. The positive magnetic field values are plotted to the right and to the left from the zero mark. The PL intensity is given by color. The arrows with numbers match the PL lines marked in Fig.~\ref{Flo:figure1}.}
\label{Flo:figure2}
\end{figure}

Evolution of the circularly polarized PL spectra in the magnetic field up to 3\,T is shown in figure~\ref{Flo:figure2}. The left and the right part of plot represent the $\sigma^+$ and $\sigma^-$ polarizations, respectively. Lines formed by the PL peaks are curved upwards due to a diamagnetic shift. A difference in the line shifts for $\sigma^+$ and $\sigma^-$ polarizations is the clear indication of the line splitting discussed below.

Besides the exciton lines, several weaker spectral lines are observed, which start from energy~$E\approx 1.489$\,eV and demonstrate almost linear behavior. 
These are the states originated from the excited states of exciton. Similar to the 2s, 3s, etc. hydrogen states, they have greater mean electron-hole distance as compared to the ground state. 
Hence they readily reach the so-called diamagnetic exciton limit~(the Loudon criterion~\cite{Elliott-Loudon}) at relatively low magnetic field. In the diamagnetic exciton, the electron and the hole are mainly confined by the magnetic field and only a weak confinement along the magnetic field is caused by the Coulomb attraction. In the limit of high magnetic field, the carriers occupy Landau levels, and the transition energy exhibits linear dependence on the magnetic field.
The energy interval between the ground exciton state and the point at zero Tesla where Landau levels meet is the experimentally observed 1s-2s distance. Accurate data processing gives value of 3.2$\pm$0.2\,meV. This value is in good agreement with the previous experimental observations and theoretical studies for bulk GaAs~\cite{Baldereschi.Lipari, binding_energy_exp}.

\begin{figure}
\includegraphics[scale=1]{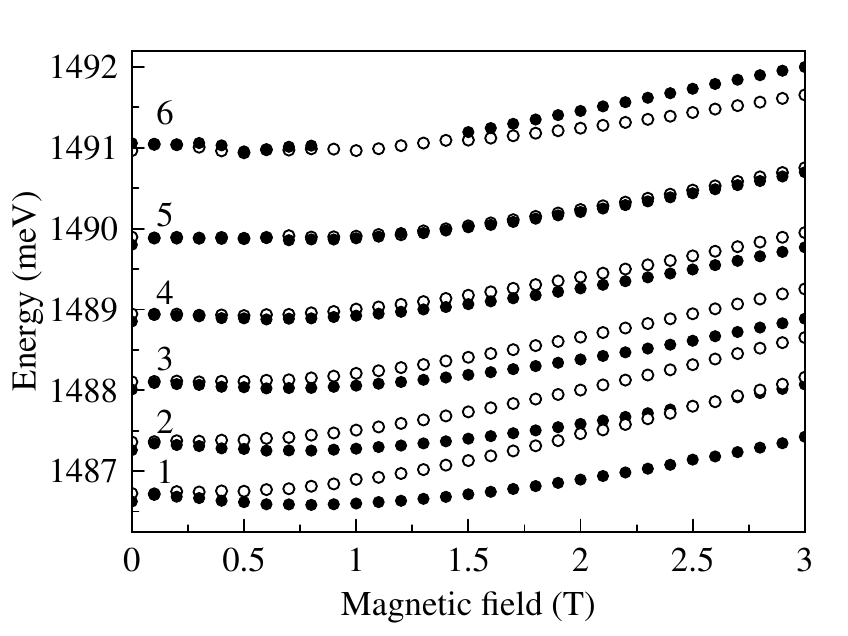}
\caption{Position of spectral peaks versus magnetic field obtained from the data shown in Fig.~\ref{Flo:figure2}. The filled and and empty circles are the different circular polarizations.}
\label{Flo:figure3}
\end{figure}

The spectral positions of the quantum-confined exciton states in magnetic field obtained by the Lorentzian fits of the PL spectra (see example in Fig.~\ref{Flo:figure1}) are plotted in figure~\ref{Flo:figure3}. The PL in opposite circular polarizations exhibits clear splitting, which decreases with the exciton state number increasing and even becomes inverse for the 6-th state. The fifth state here is of particular interest as its total magnetic momentum appears to be zero, as experiment shows. In the next sections we focus on the splitting behavior and develop a theory to explain this phenomenon.

\section{Theory}
\label{theory}

We consider an exciton as a Coulomb-interacting electron-hole pair. The conduction band is two-fold degenerate due to the 1/2 electron spin. The valence band in a semiconductor of the GaAs type has a four-fold degenerate structure, which is described by the Luttinger Hamiltonian~\cite{luttinger}. The exciton Hamiltonian can be written in the basis of eigenstates of the z-projection of the hole angular momentum operator, $\hat{J}_z$:

\begin{widetext}
\begin{equation}
{\scriptsize 
\begin{aligned}
&\hat{H}=\frac{\hat{k_e}^2}{2 m_e}I+\frac{\hat{k_h}^2 \gamma_1}{2 m_0}I
+\frac{\left(\hat{k}^2_x+\hat{k}^2_y-2\hat{k}^2_z \right)\gamma_2}{2 m_0}
\begin{pmatrix}
 1 & 0 & 0 & 0\\
 0 & -1 & 0 & 0\\
 0 & 0 & -1 & 0\\
 0 & 0 & 0 & 1
\end{pmatrix}
-\frac{e^2}{\varepsilon r}I+V(z_e,z_h)I+\mu_B g_{h} B \hat{J}_z+\mu_B g_{e} B s_z I+\\
&
+\frac{\sqrt{3}\left(\hat{k}^2_y-\hat{k}^2_x \right)\gamma_2}{2 m_0}
\begin{pmatrix}
 0 & 0 & 1 & 0\\
 0 & 0 & 0 & 1\\
 1 & 0 & 0 & 0\\
 0 & 1 & 0 & 0
\end{pmatrix}
-\frac{\sqrt{3} \gamma_3}{m_0}
\begin{pmatrix}
 0 & i\{\hat{k}_x,\hat{k}_z\}+\{\hat{k}_y,\hat{k}_z\} & i\{\hat{k}_x,\hat{k}_y\} & 0\\
 -i\{\hat{k}_x,\hat{k}_z\}+\{\hat{k}_y,\hat{k}_z\} & 0 & 0 & i\{\hat{k}_x,\hat{k}_y\}\\
 -i\{\hat{k}_x,\hat{k}_y\} & 0 & 0 & -i\{\hat{k}_x,\hat{k}_z\}-\{\hat{k}_y,\hat{k}_z\}\\
 0 & -i\{\hat{k}_x,\hat{k}_y\} & i\{\hat{k}_x,\hat{k}_z\}-\{\hat{k}_y,\hat{k}_z\} & 0
\end{pmatrix}
\end{aligned}
\label{full}
}
\end{equation}
\end{widetext}
In this expression, $m_e$ is the electron effective mass, $m_0$ is the free electron mass, $\hat{k}_{e}$ ($\hat{k}_{h}$) is the momentum operator of electron (hole), and $I$ is the unity $4\times4$ matrix. Operators $\hat{k}_x$, $\hat{k}_y$ and $\hat{k}_z$ are the components of the hole momentum operator, quantities $\gamma_1$, $\gamma_2$ and $\gamma_3$ are the Luttinger parameters, $\varepsilon$ is the dielectric constant of semiconductor, $r$ is the relative electron-hole distance and $e$ is the electron charge. Function $V(z_e,z_h)$ stands for the square QW potential. The last two terms in the first line describe the ordinary Zeeman effect for a hole and for an electron. Quantities $g_h$ and $g_e$ are the bare hole and electron $g$-factors, respectively. 
The diagonal matrix $\hat{J}_z=(+3/2,+1/2,-1/2,-3/2)$ describes the $z$-projection of the hole angular momentum. The $z$-projection of electron spin is described by the value $s_z$. The Zeeman terms are written for magnetic field $B$ applied along the z-direction. Figure brackets, $\{\hat{k}_\alpha,\hat{k}_\beta\}$, stand for the anti-commutator of operators:
\begin{equation}
\{\hat{k}_\alpha,\hat{k}_\beta\}=\frac{\hat{k}_\alpha\hat{k}_\beta+\hat{k}_\beta\hat{k}_\alpha}{2}
\end{equation}
In the presence of magnetic field $\mathbf{B}$, operators $\hat{k}_{e,h}$ should be generalized, using the symmetric gauge:
\begin{equation}
\hat{k}_{e,h}=-i\hbar\nabla_{e,h}\pm\frac{e}{2c}\left[\mathbf{B}\times \mathbf{r}_{e,h}\right]
\label{k}
\end{equation} 
where $\mathbf{r}_{e}$($\mathbf{r}_{h}$) is the electron (hole) radius vector.

Expression~(\ref{full}) is the Hamiltonian for an exciton in a QW heterostructure consisting of semiconductor layers of cubic symmetry. The Schr\"{o}dinger equation with Hamiltonian~(\ref{full}) can not be solved analytically in general case. In the case of a bulk semiconductor, the valence-band-describing terms can be rearranged in two matrices: the diagonal one and small addition with both diagonal and non-diagonal elements, which can be treated as perturbation~\cite{Baldereschi.Lipari}. The unperturbed Hamiltonian is decomposed into four independent Hamiltonians. The Schr\"{o}dinger equation with  these Hamiltonians describes separately the internal electron-hole motion and the center-of-mass (CM) motion of the hh- and lh-excitons. Resulting eigenfunctions are the plain waves for the CM-motion and the hydrogen-like functions for the relative electron-hole motion.

For an exciton in a QW, similar separation of variables is impossible even ignoring the hh-lh coupling. In particular, the introduction of CM coordinates does not separate variables along the z-axis. For a QW of an intermediate width, the consideration of a QW potential as a perturbation to the Coulomb potential leads to unacceptable controversies. Therefore, we have to use a numerical procedure to solve the six-dimensional problem for an electron and a hole interacting by the Coulomb potential and confined in a finite-depth QW of an intermediate width.

The peculiarities of studied heterostructures further complicates the problem. First, the lattice constants of InAs and GaAs differ, therefore the InGaAs/GaAs QW is strained. The strain induces a hh-lh splitting that results in a decrease of the hh-lh coupling compared to the unstrained material. Second, a segregation of indium atoms during the growth process changes the average width of the QW and breaks the presumed rectangular profile of the QW potential~\cite{Muraki, arXiv-asymmetric-QW}. 
We account for this effect choosing an appropriate QW width to get good correspondence of the calculated exciton energy spectrum with that obtained experimentally. 

We propose the numerical solution of the problem in two steps. First, we solve the Schr\"{o}dinger equation with the basic Hamiltonian, which is the first line of Hamiltonian~(\ref{full}). Then we use the obtained wave functions as a constrained basis to compose a Hamiltonian matrix and diagonalize it.

At the first step, we make use of cylindrical symmetry of the problem with the basic Hamiltonian and divide it into the problems of smaller dimensionality. In particular, the  Schr\"{o}dinger equations for excitons with the heavy and light holes can be solved separately. 
The movement of exciton as a whole along the QW layer (the xy-plane) can be separated from the relative electron-hole motion in this plane. The corresponding Schr\"{o}dinger equation is readily solved with plane waves as wave functions describing the exciton CM motion in the xy-plane.  Introducing the cylindrical coordinates $\rho$ and $\varphi$ for the xy-plain relative motion, we obtain the analytical  dependence of exciton wave function on $\varphi$ as $e^{-i k_\varphi \varphi}$. Here $k_\varphi$ is the $z$-projection of the exciton orbital momentum, which is conserved due to the basic Hamiltonian symmetry.

At this point we have got the analytical solution for the exciton CM motion in the xy-plane and for the orbital electron-hole motion in this plane. The rest of the wave function depending on $\rho$, $z_e$, and $z_h$ coordinates should be obtained numerically solving the three-dimensional eigenvalue problem. When a magnetic field is applied along the z-axis, the conserved cylindrical symmetry allows one similar wave function factorization~\cite{Gorjkov}, which we discuss in the next section. 
 
At the second step, we form a matrix of total Hamiltonian~(\ref{full}) using the obtained wave functions of the basic Hamiltonian, $\psi_n$, as a basis. Elements of the Hamiltonian matrix, $\left<\psi_n\right|\hat{H}\left|\psi_m\right>$, are calculated numerically and analytically when possible. Diagonalization of the matrix of the total Hamiltonian~(\ref{full}) with generalized operators~(\ref{k}) allows one to obtain the Zeeman splittings for a given value of magnetic field. For each value of magnetic field the basis and all the matrix elements have to be recalculated as the magnetic field affects the wave functions in the basis. On the other hand, the magnetic field makes the spectrum sparse and thus decreases the density of states in the range of interest. This sufficiently simplifies the numerical calculations.

\section{Modeling\label{modeling}}

\subsection{Step 1: obtaining of finite basis}

In this subsection we discuss the numerical computation of wave functions of a basic Hamiltonian [the first line in Eq.~(\ref{full})]. The basic Hamiltonian reads:

\begin{eqnarray}
\label{basic_H}
\hat{H}_{b}&=&\frac{\hat{k}_e^2}{2  m_e} + \frac{\left(\hat{k}_{hx}^2+\hat{k}_{hy}^2\right)\left(\gamma_1\pm\gamma_2\right)}{2 m_0}\nonumber\\
& & + \frac{\hat{k}_{hz}^2\left(\gamma_1\mp 2\gamma_2\right)}{2 m_0} + V(z_e,z_h) - \frac{e^2}{\varepsilon r},
\end{eqnarray}
the upper (lower) sign here corresponds to the hh (lh)-exciton. The QW potential is:
\begin{eqnarray}
V(z_e,z_h)&=&\left[h(a-z_e)+h(z_e-b)\right]V_e \nonumber\\
&+&\left[h(a-z_h)+h(z_h-b)\right]V_h,
\label{sqcup}
\end{eqnarray}
where $h(x)$ is the Heaviside function, $z_{e,h}$ and $V_{e,h}$ are the $z$-coordinates  and the QW depths for electron and hole, respectively. In the calculations described below, we assume that $V_{e} = 2 V_{h}$, which is typical ratio for GaAs/InGaAs/GaAs QWs with small In content. Heterostructures InGaAs/GaAs are strained  due to the lattice constants mismatch. The strain results in the hh-lh splitting, which decreases the depth of potential well for the light-hole. We take into account this splitting as it is described in Sect.~\ref{sample-P554}.

Schr\"{o}dinger equation with Hamiltonian~(\ref{basic_H}) can be solved independently for the hh- and lh-excitons. To simplify Eq.~(\ref{basic_H}), we introduce effective masses for the heavy and the light holes:
\begin{equation}
m_{hxy}=\frac{m_0}{\gamma_1 \pm \gamma_2}\quad m_{hz}=\frac{m_0}{\gamma_1 \mp2 \gamma_2}
\end{equation}  
The upper (lower) signs are again used for the heavy (light) holes. 

To separate the relative motion of an electron and a hole in the exciton from the motion of exciton as a whole, a conventional definition of the center-of-mass (CM) and relative coordinates in $xy$-plane is used:
\begin{equation}
\begin{aligned}[r|r]
X&=\frac{m_e x_e+m_{hxy} x_h}{m_e+m_{hxy}}\quad & Y=&\frac{m_e y_e+m_{hxy} y_h}{m_e+m_{hxy}}\\
x&=x_e-x_h=\rho \cos\,\varphi & y=&y_e-y_h=\rho \sin\,\varphi\\
\label{CMcoord}
\end{aligned}
\end{equation}

With introduced polar coordinates for the $xy$-plane relative motion, the basic Hamiltonian has a form:
\begin{eqnarray}
\hat{H}_{b}&= &\frac{\hbar^2 \left(K_X^2+K_Y^2\right)}{2 \left(m_e+m_{hxy}\right)}-\frac{\hbar^2}{2 \mu_{xy}}\left[\frac{1}{\rho}\frac{\partial}{\partial \rho}\left(\rho\frac{\partial}{\partial \rho}\right)-\frac{k_\varphi^2}{\rho^2}\right]\nonumber\\
 &-&\frac{\hbar^2}{2 m_e}\frac{\partial^2}{\partial z_e^2}-\frac{\hbar^2}{2 m_{hz}}\frac{\partial^2}{\partial z_h^2}\nonumber\\
&-&\frac{e^2}{\varepsilon \sqrt{\left(z_e-z_h\right)^2+\rho^2}}+V(z_e, z_h).
\label{finalBasic}
\end{eqnarray}
Here $\mu_{xy}=(m^{-1}_e+m^{-1}_{hxy})^{-1}$ is the reduced exciton mass in the $xy$-plane.  The corresponding wave function has a partially analytical form:
\begin{equation}
\psi(X,Y,z_e,z_h,\rho,\varphi)=e^{iK_XX}e^{iK_YY}e^{i k_\varphi \varphi}\frac{\psi(z_e,z_h,\rho)}{\rho}. 
\label{psiXY}
\end{equation}
Here we introduce denominator $\rho$ for the convenience of the numerical solution. With wave function in form of~(\ref{psiXY}) we arrive to the following three-dimensional problem, which requires numerical calculations:
\begin{eqnarray}
&\left[ -\frac{\hbar^2}{2 m_e}\frac{\partial^2}{\partial z_e^2}-\frac{\hbar^2}{2 m_{hz}}\frac{\partial^2}{\partial z_h^2}-\frac{\hbar^2}{2 \mu_{xy}}\left(\frac{\partial^2}{\partial \rho^2}-\frac{1}{\rho}\frac{\partial}{\partial \rho}+\frac{1-k_\varphi^2}{\rho^2}\right) \right.\nonumber\\
&\left.-\frac{e^2}{\varepsilon \sqrt{\left(z_e-z_h\right)^2+\rho^2}}+V \right] \psi(z_e,z_h,\rho)=E\psi(z_e,z_h,\rho).
\label{maincylinder}
\end{eqnarray}

The described above coordinate separation is not exact in the presence of magnetic field.
To take into account the magnetic field, one should use a generalized momentum operator. We restrict our treatment to the Faraday geometry case. In that extent, the momentum operator  is generalized according to expression~(\ref{k}). 
\begin{equation}
\left\{
\begin{aligned}
\hat{k_x}&=&-i\hbar\frac{\partial}{\partial x}&\mp\frac{e}{2c}B y,\\
\hat{k_y}&=&-i\hbar\frac{\partial}{\partial y}&\pm\frac{e}{2c}B x,\\
\hat{k_z}&=&-i\hbar\frac{\partial}{\partial z}&.
\end{aligned}
\right.
\label{kparticular}
\end{equation}
The upper (lower) sign here corresponds to electron (hole). Gorjkov and Dzjaloshinskiy~\cite{Gorjkov} have showed that, in the exciton Hamiltonian with momentum operators in form of~(\ref{kparticular}), one can separate the CM coordinates with the wave function in the form of ansatz:
\begin{equation}
	\psi=\exp\left[i\frac{e B}{2 c \hbar}\left(xY-yX\right)\right]\psi\left(z_e,z_h,\rho,\varphi\right).
\label{ansatz}
\end{equation}
At this point we assume that the CM kinetic energy in the $xy$-plane is zero ($K_X=K_Y=0$). The basic Hamiltonian~(\ref{finalBasic}) with suggested ansatz acquires the following form:
\begin{eqnarray}
\begin{aligned}
&\hat{H}_{b}(B)=\hat{H}_{b}+\frac{\rho^2}{2\mu_{xy}}\left(\frac{e B}{2 c}\right)^2 \\
&-i\frac{e\hbar B}{2c}\left(\frac{m_h-m_e}{M\mu_{xy}}\right)\frac{\partial}{\partial \varphi}+g_e \mu_B \sigma_z B+g_h \mu_B J_z B.
\label{maincylindermagnetic}
\end{aligned}
\end{eqnarray}
Angular dependency of the wave function is still valid for the Hamiltonian~(\ref{maincylindermagnetic}). The net exciton wave function in the presence of magnetic field yields:
\begin{eqnarray}
\begin{aligned}
\psi_B(X,Y,\rho,\varphi,z_e,&z_h)_{ljk_\varphi}=&\\
\exp\left[{i\frac{e B \rho}{2 c \hbar}\left(Y\cos{\varphi}-X\sin{\varphi}\right)}\right]
 &e^{i k_\varphi \varphi}\psi_B(z_e,z_h,\rho)_{lk_\varphi j}.
\end{aligned}
\label{wfB}
\end{eqnarray}
Here  $j$ ($j = \pm1/2$ and $\pm3/2$) indicates certain $z$-projection of the hole angular momentum and  $k_{\varphi} = 0, \pm 1, \pm 2, \ldots$ indicates certain $z$-projection of the exciton orbital momentum. We use index $l = 0,1,2,\ldots$ to numerate different exciton states for given value of $j$ and $k_\varphi$. For each value of $j$ and $k_\varphi$, we obtain a Hamiltonian for the three-dimensional problem, similar to problem~(\ref{maincylinder}),  which takes the form: 
\begin{eqnarray}
\label{basicmagn} 
\hat{H}_{b}(B)^{3D} &=&-\frac{\hbar^2}{2 m_e}\frac{\partial^2}{\partial z_e^2}-\frac{\hbar^2}{2 m_{hz}}\frac{\partial^2}{\partial z_h^2}\\ \nonumber
&-&\frac{\hbar^2}{2 \mu_{xy}}\left(\frac{\partial^2}{\partial \rho^2}-\frac{1}{\rho}\frac{\partial}{\partial \rho}+\frac{1-k_\varphi^2}{\rho^2}\right) \\ \nonumber
&-&\frac{e^2}{\varepsilon \sqrt{\left(z_e-z_h\right)^2+\rho^2}}+V+\frac{\rho^2}{2\mu_{xy}}\left(\frac{e B}{2 c}\right)^2.
\end{eqnarray}
The eigenproblem with this operator is solved numerically, separately for the heavy-hole and light-hole excitons ($j=\pm3/2$ and $j = \pm1/2$, respectively). The value of magnetic field is set before the numerical procedure is performed.

The wave function set~(\ref{wfB}) forms a complete orthonormal system of functions with magnetic field as an extra parameter. Strictly speaking, this set is the infinite system of exciton wave functions in a QW. However we are interested in the several lowest exciton states, which are observed in the PL experiments (see Fig.~\ref{Flo:figure1}). Therefore we restrict the basis to the observed $s$-like states with $k_{\varphi}=0$ and to $p$-like and $d$-like states, which are significantly coupled with the $s$-like states (see next subsection).

We use the obtained basis to build a matrix of total Hamiltonian~(\ref{full}). The diagonal matrix elements follow directly from Hamiltonian~(\ref{maincylindermagnetic}):
\begin{equation}
\begin{aligned}
H&_{nmlk_\varphi}^{hh}=E_{lk_\varphi}^{hh}+\frac{e \hbar B}{2 c }\left(\frac{m_h-m_e}{M\mu_{xy}}\right) k_\varphi \pm\\
&\pm\frac{3}{2}g_h \mu_B B\mp\frac{1}{2}g_e \mu_B B,\\
H&_{nmlk_\varphi}^{lh}=E_{lk_\varphi}^{lh}+\frac{e \hbar B}{2 c }\left(\frac{m_h-m_e}{M\mu_{xy}}\right) k_\varphi \pm\\
&\pm\frac{1}{2}g_h \mu_B B\mp\frac{1}{2}g_e \mu_B B.
\label{diagmatrix}
\end{aligned}
\end{equation}
Here $E_{lk_\varphi}^{hh,lh}$ are the eigenvalues of operator~(\ref{basicmagn}). 
The second terms in these expressions describe the interaction of excitonic orbital momentum with the magnetic field. The last two terms describe the exciton Zeeman splitting related to the electron and hole magnetic momenta.

The electron Zeeman term in Eq.~(\ref{diagmatrix}) has an opposite sign compared to that for hole term, because the angular momentum of the optically active (bright) hh-exciton is the difference of the electron spin and the hole angular momentum. The electron and hole $g$-factors, $g_e$ and $g_h$, are changed due to the interband mixing. The electron $g$-factor in III-V semiconductors can be obtained as~\cite{Roth}.
\begin{equation}
g_e=2-\frac{2E_p\Delta_{so}}{3E_g(E_g+\Delta_{so})}
\end{equation}
Here $E_p$ is the optical matrix element, $\Delta_{so}$ is the spin-orbit band offset, and $E_g$ is the band gap. For the GaAs, the electron $g$-factor, $g_e=-0.44$~\cite{Yugova-PRB2007}. The hole $g$-factors, which we use for the calculations, are connected to the Luttinger parameter $\kappa$ as~\cite{luttinger}:
\begin{equation}
\kappa=-\frac{g_{hh}}{6}=-\frac{g_{lh}}{2}
\end{equation}
here $g_{hh,lh}$ are the hole $g$-factors ($\kappa=1.2$ for GaAs).

\subsection{Step 2: the total Hamiltonian diagonalization in finite basis}

The first step provides eigenfunctions for the basic Hamiltonian. In the InGaAs/GaAs QWs (that we are interested in here), the strain-induced valence-band splitting partially supresses the hh-lh coupling. Therefore the eigenfunctions of the basic Hamiltonian are good approximation to the eigenfunctions of the system without the magnetic field.

The second step accounts for the hh-lh coupling induced by the magnetic field. We compose a suitable basis to describe the bright hh-exciton states, which we observe in the experiment. In this basis, we build a matrix of the total Hamiltonian~(\ref{full}). The matrix consists of matrix elements $\{H_{\eta\prime\eta}\}$ defined as:
\begin{equation}
H_{\eta^\prime\eta}=\left<\psi_{B\eta^\prime}\right|\hat{H}\left|\psi_{B\eta}\right>.
\label{Hmat}
\end{equation}
Here $\eta$ and $\eta^\prime$ stand for $ljk_\varphi$ and $l^\prime j^\prime k^\prime_\varphi$, respectively, $\{\psi_{B\eta}\}$ is the restricted basis formed from set~(\ref{wfB}). 

The restricted basis should include the optically active states and all the eigenfunctions $\psi_B$ admixed by the nondiagonal terms in Hamiltonian $\hat{H}$. The nondiagonal terms couple hh-exciton states to the lh-exciton ones only. The exciton wave functions in form of~(\ref{wfB}) have different $X$ and $Y$ coordinates for hh- and lh-exciton states. The used ansatz, however, allows one to ignore this fact in the calculation of matrix element as discussed in the Appendix. The ansatz provides simplification of coupling operators as well.  The simplified operators have the following structure:
\begin{eqnarray}
\hat{k}_y^2-\hat{k}_x^2&=&2\sin{2\varphi}\,L_{\partial_\rho,\partial_\varphi, \rho}+\cos{2\varphi}\,L_{\partial_\rho,\partial_\varphi, \rho}^\prime
\label{ndperturbation1}\\
\{\hat{k}_x,\hat{k}_z\}&=&-\sin{\varphi}\,L_{\partial_\varphi, \rho}\partial_z+\cos{\varphi}\,\partial_\rho \partial_z
\label{ndperturbation2}\\
\{\hat{k}_y,\hat{k}_z\}&=&\sin{\varphi}\,\partial_\rho \partial_z+\cos{\varphi}\,L_{\partial_\varphi, \rho}\partial_z
\label{ndperturbation3}\\
\{\hat{k}_x,\hat{k}_y\}&=&-\frac{1}{2}\sin{2\varphi}\,L_{\partial_\rho,\partial_\varphi, \rho}^\prime+\cos{2\varphi}\,L_{\partial_\rho,\partial_\varphi, \rho}
\label{ndperturbation4}
\end{eqnarray}
In these expressions, $\partial_\alpha$ is the partial derivative with respect to the $\alpha$ variable. Quantities $L_{\alpha,\beta}$ stand for combination of $\alpha$ and $\beta$ operators explained in the Appendix. Matrix elements of coupling operators are nonzero if $k_\varphi$ of two states differ by 1 for Eqs.~(\ref{ndperturbation2}) and~(\ref{ndperturbation3}), and by 2 for Eqs.~(\ref{ndperturbation1}) and~(\ref{ndperturbation4}). We therefore consider 5 orbital momentum projections ($k_\varphi=0,\pm1,\pm2$) to describe the magnetic-field-induced admixture of the light-hole exciton states to the observed heavy-hole exciton states with $k_\varphi^\prime=0$.

These simple selection rules can be refined. The nondiagonal matrix elements of $H$ are the linear combinations of matrix elements of coupling operators~(\ref{ndperturbation1})--(\ref{ndperturbation4}). In the notations presented above, the coupling matrix elements are proportional to:

\begin{eqnarray}
	H_{l^\prime\frac{3}{2}k_\varphi^\prime l\frac{1}{2}k_\varphi} & \propto & \left(e^{-i\varphi}\left[i\partial_\rho \partial_z+L_{\partial_\varphi, \rho}\partial_z\right]\right)_{l^\prime\frac{3}{2}k_\varphi^\prime l\frac{1}{2}k_\varphi}\label{H12}\\
	H_{l^\prime-\frac{3}{2}k_\varphi^\prime l-\frac{1}{2}k_\varphi} & \propto & \left(e^{i\varphi}\left[i\partial_\rho \partial_z-L_{\partial_\varphi, \rho}\partial_z\right]\right)_{l^\prime-\frac{3}{2}k_\varphi^\prime l-\frac{1}{2}k_\varphi}\label{H43}
\end{eqnarray}
\begin{widetext}	
\begin{eqnarray}
H&&_{l^\prime\frac{3}{2}k_\varphi^\prime l-\frac{1}{2}k_\varphi} \propto \left(\gamma_2 e^{2 i \varphi}\left[L_{\partial_\rho,\partial_\varphi, \rho}^\prime-2 i L_{\partial_\rho,\partial_\varphi, \rho}\right]\right.+\left.i\left(\gamma_3-\gamma_2\right)\left[\sin{2\varphi}\,L_{\partial_\rho,\partial_\varphi, \rho}^\prime-2\cos{2\varphi}\,L_{\partial_\rho,\partial_\varphi, \rho}\right]\right)_{l^\prime\frac{3}{2}k_\varphi^\prime l-\frac{1}{2}k_\varphi} \label{H13} \\
H&&_{l^\prime-\frac{3}{2}k_\varphi^\prime l\frac{1}{2}k_\varphi} \propto \left(\gamma_2 e^{-2 i \varphi}\left[L_{\partial_\rho,\partial_\varphi, \rho}^\prime + 2 i L_{\partial_\rho,\partial_\varphi, \rho}\right]\right.-\left.i\left(\gamma_3-\gamma_2\right)\left[\sin{2\varphi}\,L_{\partial_\rho,\partial_\varphi, \rho}^\prime-2\cos{2\varphi}\,L_{\partial_\rho,\partial_\varphi, \rho}\right]\right)_{l^\prime-\frac{3}{2}k_\varphi^\prime l\frac{1}{2}k_\varphi} \label{H42}
\end{eqnarray}
\end{widetext}
Expression~(\ref{H12}) is nonzero for $\left(k_\varphi,k_\varphi^\prime\right)=\left(1,0\right)$, while expression~(\ref{H43}) is nonzero for $\left(k_\varphi,k_\varphi^\prime\right)=\left(-1,0\right)$. Matrix elements~(\ref{H13}) and~(\ref{H42}) are nonzero for $\left(k_\varphi,k_\varphi^\prime\right)=\left(\pm2,0\right)$. We however found that the second term in these expressions gives rise to the considerably smaller contribution than the first term that is the major effect of coupling with $d$-like states comes from the first terms.  It is non-zero for $\left(k_{\varphi},k_{\varphi}^\prime\right)=\left( -2,0 \right)$ for matrix element~(\ref{H13}) and $\left(k_{\varphi} k_{\varphi}^\prime\right) = \left(2, 0\right)$ for~(\ref{H42}).
 
 \subsection{Numerical results for sample P554}
 \label{sample-P554}
 
 The exciton states are characterized with projections of three angular momenta (on the  magnetic field direction): electron spin, hole spin, and exciton orbital momentum. Table~\ref{BasisTable} represents the basis used to build the Hamiltonian matrix $H$ for calculations of $g$-factors for exciton states in the 87-nm QW in sample P554. The number of states in each group was determined studying the saturation of the effect of this group on the observed states (see Fig.~\ref{Flo:figure7} in the Appendix). The electron spin projection for the $s$-like states is taken so that the states would be bright. For other states, the electron spin coincides with the spin projection of $s$-like state it couples with. The coupling selection rules are denoted in the table by the brackets on the right side.

\begin{table}[htbp!]
\begin{ruledtabular}
\caption{Number of states in the restricted basis, $N$, used in the second step of calculations. The states are grouped by angular momenta projections. Symbols $\rceil$ and $\rfloor$ in each column indicate the coupling states.\label{BasisTable}}
\begin{tabular}{ c c c c c c c c c}
$\left|s,j,k_\varphi\right>$ & N & & & & & &\\
\hline
$\left|-\frac{1}{2},\frac{3}{2},0\right>$ & 5 &  $_\rceil$ & $_\rceil$ & $_\rceil$ & & \\
$\left|\frac{1}{2},-\frac{3}{2},0\right>$ & 5 & & & & $_\rceil$ & $_\rceil$ & $_\rceil$\\
$\left|-\frac{1}{2},\frac{1}{2},1\right>$ & 400 & $^\rfloor$ & & & & & \\
$\left|\frac{1}{2},-\frac{1}{2},-1\right>$ & 400 & & & & $^\rfloor$ & & \\
$\left|-\frac{1}{2},-\frac{1}{2},2\right>$ & 200 & & $^\rfloor$ & & & & \\
$\left|\frac{1}{2},\frac{1}{2},2\right>$ & 200 & & & & & $^\rfloor$ & \\
$\left|-\frac{1}{2},-\frac{1}{2},-2\right>$ & 200 & & & $^\rfloor$ & & & \\
$\left|\frac{1}{2},\frac{1}{2},-2\right>$ & 200 & & & & & & $^\rfloor$\\
\end{tabular}
\end{ruledtabular}
\end{table}
 
\begin{figure}
\includegraphics[scale=1]{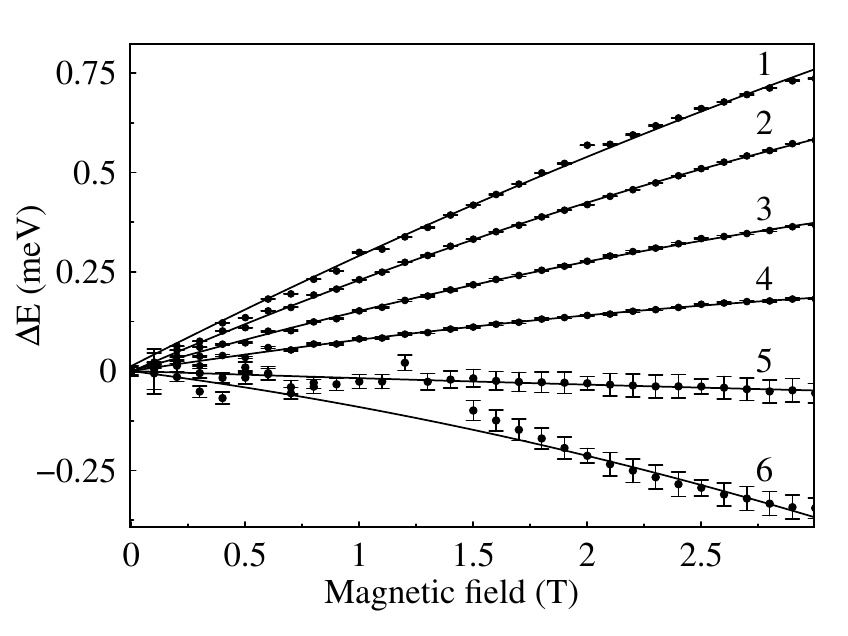}
\caption{Splittings of the observed exciton states in magnetic field for the 87-nm QW in sample P554 (black points with error bars). Numbers correspond to the exciton state numbers. The solid lines are the parabolic approximations of the splittings. For the splitting of exciton level No.~6, there are no reliable data for the magnetic field range $B = 0.8 \div 1.5$~T because of crossing of the level with the Landau levels, see Fig.~\ref{Flo:figure2}.}
\label{Flo:figure4}
\end{figure}

With this basis we calculate elements of matrix $H$ for a given magnetic field value. We then obtain eigenvalues of the matrix and extract the Zeeman splittings of the observed states. The experimentally observed Zeeman splittings are nonlinear in magnetic field as it is shown in Fig.~{\ref{Flo:figure4}}. Our modeling procedure might describe the observed nonlinearities. However, because of high complexity of the numerical procedure, we perform the computation and determine the Zeeman splitting only at the fixed magnetic field $B=1$\,T.

\begin{figure}
\includegraphics[scale=1]{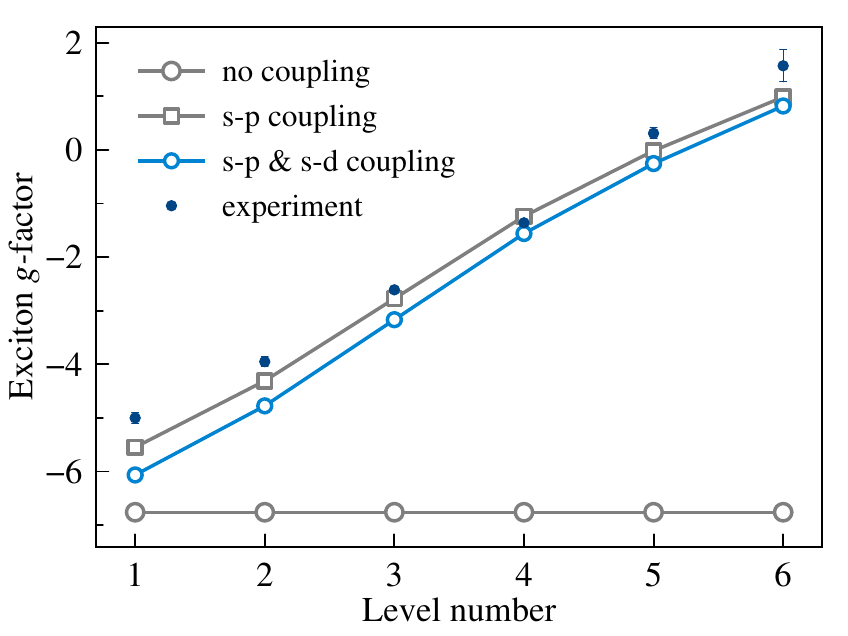}
\caption{Experimentally obtained $g$-factors of excitons at $B = 1\,$T versus level number for the 87-nm QW in sample P554 (blue points with error bars). The grey and pale blue rounds are the calculation results with no coupling and the hh-lh coupling, respectively. The contribution of s-p coupling only is shown by the blank squares.}
\label{Flo:figure5}
\end{figure}

Figure~\ref{Flo:figure5} shows the results of the calculations. In this figure, the numerically obtained values of the exciton $g$-factor for different quantum confined states are compared with those extracted from the experimental data shown in Fig.~\ref{Flo:figure3} at magnetic field $B = 1\,$T. As seen, the numerical simulation well reproduces the main experimental result. In the figure, we denote hh-lh coupling between the hh-exciton states with $k_\varphi=0$ ($s$-like states) and the lh-exciton states with $k_\varphi=\pm1$ ($p$-like states) as the $s$-$p$-coupling. The coupling with the $k_\varphi=\pm2$ ($d$-like) states is denoted as the $s$-$d$-coupling.

We found that the $s$-$p$-coupling is the main origin of the $g$-factor variation. We also found that the contribution of $p$-like states with positive momenta projections onto the magnetic field prevails over the contribution of states with opposite projections. This leads to the increase of exciton $g$-factor as shown in figure~\ref{Flo:figure5}. 

The $s$-$d$-coupling, in turn, can be expressed by two terms [see Eqs.~(\ref{H13}) and~(\ref{H42})]. The first term is greater than second one and undergoes a selection rule. It couples the $s$-like states with the positive hole spin projection to states with the negative projections of both the hole spin and the orbital momentum (and vice versa). The coupling with the positive hole spin projection is weaker. As a result, the $s$-$d$-coupling leads to the opposite effect on $g$-factor as compared to the $s$-$p$-coupling, as it is seen in figure~\ref{Flo:figure5}. There is some deviation of the calculated $g$-factors from the measured ones. It can be attributed to uncertainties of parameters $\gamma_3$ and $\kappa$ used in the calculations. We used the values of these parameters corresponding to the bulk GaAs. We intentionally avoided any variation of these parameters as their values are not reliably known as for pure InAs as for InGaAs ternary alloy. 

The sign of exciton $g$-factor requires a separate discussion. We put it negative for the lowest quantum-confined exciton states in the QW under study. However our results do not allow us to uniquely determine the sign. The sign of the hole $g$-factor and, correspondingly, of the exciton $g$-factor is extensively discussed in literature, see, e.g., Refs.~\cite{Snelling-PRB1992, Traynor-PRB1997, Kotlyar-PRB2001, Chen-APL2006, Kubisa-PRB2012, Arora-JAP2013, Jadczak-APL2014, Semina-Semicond2015, Yakovlev-PRB2011, van-Bree-PRB2016, Semina-PRB2016, Belykh-PRB2016}. However, there is no certain conclusion about the sign so far.

The results of computation shown in  figure~\ref{Flo:figure5} are obtained with no fitting parameters. The parameters needed for the computation are the QW width, the magnitude of the strain-induced hh-lh splitting, $S$,  and the material parameters defining the valence band structure and the hole $g$-factor. All the material parameters for GaAs are taken from Ref.~[\onlinecite{materials}]. The hh-lh splitting energy $S$ is taken from the PL excitation spectra (not shown here). Our model accounts for this splitting by reducing the depth of the QW for the light hole, $V_{lh}$, in expression~(\ref{sqcup}) down to $V_{lh} = (V_{hh} - S)$ with  $S=7.5$\,meV for the 87-nm QW in sample P554. These values are in a good agreement with the strain splitting dependence on the In concentration described in paper by Van de Walle~\cite{Van_de_Walle}. 
The nominal QW width predefined in the MBE growth process was 95~nm in this sample. The actual (effective) QW width in the sample under study is reduced down to 87~nm due to a gradient of the heterostructure layer thicknesses. The actual width and the segregation length ($\lambda_D = 3.75$\,nm) have been obtained by modeling of the exciton energy spectrum. In the modeling, the segregations was accounted for using the diffusion model proposed in Ref.~\cite{Muraki}. Details of the exciton spectrum modeling are described in Ref.~\cite{arXiv-asymmetric-QW}. 

In our computations, we used a grid of $50\times50\times400$ points along the $z_e$, $z_h$, and $\rho$ directions, respectively, in area $120$~nm$\times120$~nm$\times800$~nm. The boundary conditions suggest eigenfunctions to be zero on the area boundaries. The computation was processed by the Arnoldi algorithm realized on a personal computer. Additional detail of the computations can be found in Ref.~\cite{2015arXiv150800480K}. Result of the computations include two sets of eigenfunctions and eigenvalues for the hh- and lh-excitons.

\section{Narrow quantum wells}

\begin{figure}
\includegraphics[scale=1]{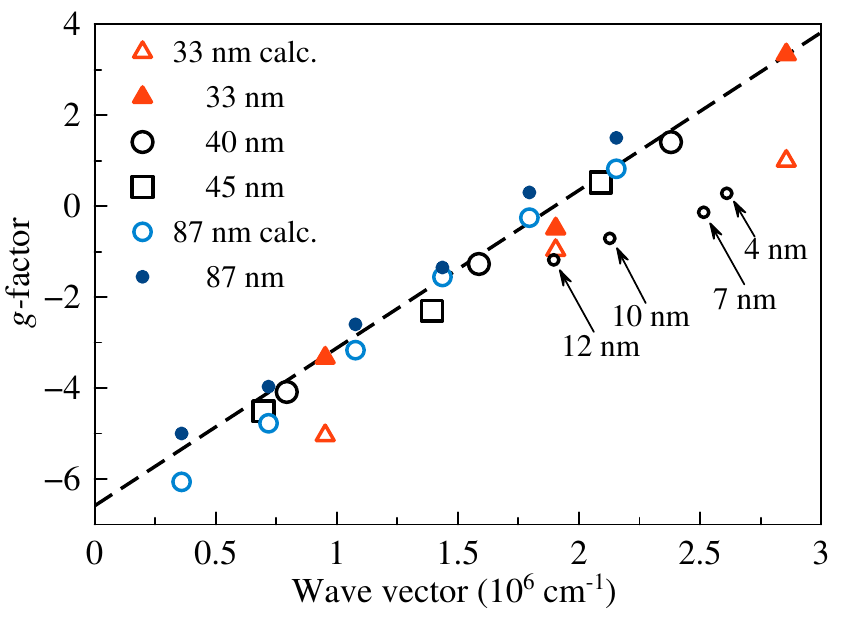}
\caption{$g$-factors of excitons versus effective hole wave vector $k_z^*$. For QWs of the 87, 45, 40, and 33-nm width, the $g$-factors of the ground and excited states are shown. Red blank triangles and pale blue circles show the results of numerical modeling of exciton $g$-factors for the 33-nm and 87-nm QWs respectively. Dashed line is the fit by linear dependence $g_{\text{ex}}=\varkappa k^*_z+g_0$ with parameters: $\varkappa=(3.5\pm0.1)\times10^{-6}\,$cm, $g_0=-6.6\pm0.2$.}
\label{Flo:figure6}
\end{figure}

The observed phenomenon of the large difference of Zeeman splittings of different exciton states is not unique property of the 87-nm QW discussed above. In this section, we demonstrate a generality of this effect by the experimental study of a number of heterostructures with QWs of different widths. We have studied a set of three high-quality InGaAs/GaAs QWs of the 30, 36 and 41\,nm widths. The actual width determined by the modeling of the exciton spectra are found to  be 10\% greater. The indium diffusion length $\lambda_D=2$~nm in this structure. Besides, four narrow QWs of the 12, 10, 7, and 4\,nm widths were studied. In the QWs, which width is of about the exciton-Bohr diameter (33~nm), three quantum confined states are observed, while, for the 12\,nm-wide and narrower QWs, only the ground state is observed. Studying polarized photoluminescence from these samples, we obtained various values of $g$-factor in the -4$\div$4 range as it is shown in Fig.~\ref{Flo:figure6}.

For the 33-nm QW, we have performed similar theoretical analysis of the $g$-factor behavior for the quantum-confined exciton states. Results of the analysis are shown in Fig.~\ref{Flo:figure6} by the red blank triangles. As seen, the theoretically obtained $g$-factors correspond to the experimentally found ones shown by the red empty triangles. Some deviation between theory and experiment is possibly related to the different values of $\gamma_3$ and $\kappa$ in the In$_{0.05}$Ga$_{0.95}$As QW than those of GaAs used in the modeling. 

An analysis shows that there is no regular dependence of $g$-factors on the exciton transition energy for different QWs (not shown here). This is in a drastic contrast to the behavior of electron $g$-factor, which is monotonically changed with the transition energy~\cite{Yugova-PRB2007}. At the same time, the $g$-factors obtained for different exciton transitions in one QW monotonically rise with the exciton state number (see Fig.~\ref{Flo:figure5}). This is an indication that some regular dependence of $g$-factors on an effective wave vector may take place.

As we already discussed above, the $g$-factor variation is described by the $\{\pm i\hat{k}_x\pm\hat{k}_y,\hat{k}_z\}$ operators coupling the hh- and lh-excitons. Therefore it is reasonable to consider the $g$-factor versus some effective wave vector of the heavy hole. 
Proper definition of the wave vector is problematic due to the Coulomb electron-hole interaction in the exciton. We, therefore, suggest some ``naive'' estimate of the wave vector. Particularly, we consider the hole wave function to be approximated by functions $\cos(k_z^{*} z)$ and $\sin(k_z^{*} z)$ for the quantum confined states $n=1,3,\ldots$ and $n=2,4,\ldots$, respectively. Here $k_z^{*} = n\pi/L$ is the $z$-projection of the effective hole wave vector. We use this definition for the QWs width down to 30\,nm. 

Figure~\ref{Flo:figure6} shows the dependence of exciton $g$-factor obtained experimentally and calculated theoretically on the $k_z^{*}$ for QWs with $L_{QW} \ge 30$~nm. As seen, an universal dependence of $g$-factor on the $k_z^{*}$ is observed, which can be well approximated by a linear function within the spread of the $g$-factors. For the QWs of width $L_{QW} \le 30$~nm, the used above definition of $k_z^{*}$ is not valid anymore. The reason is that the hole wave function penetrates to the barrier layers and the effective wave vector is not determined by the real QW width. We can roughly estimate $k_z^{*}$ considering only the central part of the wave function within the QWs. Respective $k_z^{*}$ are given in Fig.~\ref{Flo:figure6} for the narrow QWs. As seen, some deviation from the linear dependence with the decrease of the QW width is observed. We assume that the main reason for this effect is the penetration of hole wave function into the barriers.

The universal character of $g$-factor renormalization shown in Fig.~\ref{Flo:figure6} is observed not only for QWs of varying widths but also for the excited exciton states in the QWs. We should note that monotonic dependence of the exciton $g$-factor on the effective wave vector has been observed previously only for the wide QWs~\cite{kochereshko.1, kochereshko.2, kochereshko.3}.

\section{Conclusion}

Our study shows that the direct calculation of Zeeman splittings of the quantum confined exciton states is the effective method to describe the evolution of exciton systems in the longitudinal magnetic field. The theoretical analysis has shown that experimentally observed large change of exciton $g$-factor with number of quantization level in the intermediate-width and narrow QWs is caused by the mixing of the heavy-hole and the light-hole exciton states. We developed a model, which takes into account all the valuable interactions in the system. Numerical simulations with no fitting parameters quantitatively reproduce experimentally observed behavior of $g$-factors for the 87-nm and 33-nm thick QWs. It is important that our model can be used to obtain $g$-factors of excitons in the QWs of arbitrary small thickness as long as the envelope function approximation is applicable. The developed approach allows one to numerically obtain the exciton wave function in a QW, which width is comparable to the exciton Bohr radius.

\begin{acknowledgments}
The authors are grateful to M.~A.~Semina, M.~M.~Glazov, M.~V.~Durnev, and I.~Y.~Gerlovin for fruitful discussions. The financial support from the Russian Ministry of Education and Science (contract No.  11.G34.31.0067), from the Russian Fond for Basic Research (RFBR, grand No.~15-52-12019 ) in the frame of ICRC TRR-160, and from the St-Petersburg State University (SPbU, grant No. 11.38.213.2014) is acknowledged. I.V.I. acknowledges RFBR for the financial support in frame of grand No.~16-02-00245-a. The authors also thank the SPbU Resource Center ``Nanophotonics'' (www.photon.spbu.ru) for the samples studied in present work.
\end{acknowledgments}

\appendix*

\section{Non-diagonal operators of Luttinger Hamiltonian in cylindrical coordinates\label{appA}}

The non-diagonal operators of Luttinger Hamiltonian are used in our consideration in cylindrical coordinates introduced by equations~(\ref{CMcoord}). In the magnetic field perpendicular to the QW plane, this operators can be expressed in a form:
\begin{widetext}
\begin{equation}
\begin{aligned}
\hat{k}_y^2-\hat{k}_x^2=
&\sin{2\varphi}\,\hbar^2\left(
\frac{2}{\rho^2}\partial_\varphi
-\frac{2}{\rho}\partial_\rho\partial_\varphi
+2i\frac{e B}{2 c \hbar}\rho\partial_\rho
\right)
+\cos{2\varphi}\,\hbar^2\left(
\partial_\rho^2
-\frac{1}{\rho^2}\partial_\varphi^2
-\frac{1}{\rho}\partial_\rho
+2i\frac{e B}{2 c \hbar}\partial_\varphi
+\left(\frac{e B}{2 c \hbar}\right)^2\rho^2
\right)
\\
\{\hat{k}_x,\hat{k}_z\}=&\cos{\varphi}\,\hbar^2 \partial_{\rho}\partial_{z_h}- \sin{\varphi}\,\hbar^2\left(i\frac{e B}{2 c \hbar}\rho +\frac{1}{\rho}\partial_{\varphi}\right)\partial_{z_h}\\
\{\hat{k}_y,\hat{k}_z\}=&\sin{\varphi}\,\hbar^2 \partial_{\rho}\partial_{z_h}+\cos{\varphi}\,\hbar^2\left(i\frac{e B}{2 c \hbar}\rho +\frac{1}{\rho}\partial_{\varphi}\right)\partial_{z_h}\\
\{\hat{k}_x,\hat{k}_y\}=
&\sin{2\varphi}\frac{1}{2}\,\hbar^2\left(
-\partial_{\rho}^2
+\frac{1}{\rho}\partial_{\rho}
+\frac{1}{\rho^2}\partial_\varphi^2
-2i\frac{e B}{2 c \hbar}\partial_{\varphi}
-\left(\frac{e B}{2 c \hbar}\right)^2\rho^2
\right)
+\cos{2\varphi}\,\hbar^2\left(
\frac{1}{\rho^2}\partial_{\varphi}
-\frac{1}{\rho}\partial_{\varphi}\partial_{\rho}
+i\frac{e B}{2 c \hbar}\rho\partial_{\rho}
\right)
\end{aligned}
\label{ndperturbationlong}
\end{equation}
\end{widetext}
Such a form duplicates the structure of expressions~(\ref{ndperturbation1}, \ref{ndperturbation2}, \ref{ndperturbation3}, \ref{ndperturbation4}) therefore it is easy to match linear operators here with those introduced above.

In section~\ref{modeling}, we noted that the CM $X$ and $Y$ coordinates for the heavy-hole and light-hole excitons are different according to the definition. This difference implies that a matrix element on a given operator $\hat{A}$ mixing the heavy-hole states with the light-hole states should be written in terms of $X_{hh,lh}$ and $Y_{hh,lh}$ coordinates:
\begin{equation}
\left<\psi(X_{lh},Y_{lh}, z_e, z_h,\rho, \varphi)\right|\hat{A}\left|\psi(X_{hh},Y_{hh}, z_e, z_h,\rho, \varphi)\right>.
\end{equation}
The integration has to be done in one of the coordinate systems involved: the heavy-hole exciton system or the light-hole one.
However, one can show using anzats~(\ref{ansatz}) that these coordinate system are equivalent for our wave functions. Indeed, the exponent in~(\ref{ansatz}), which only contains $X$ and $Y$ coordinates, yields:  
\begin{equation}
\begin{aligned}[c]
&\left(Y_{lh}\cos{\varphi}-X_{lh}\sin{\varphi}\right)=\left(Y_{hh}\cos{\varphi}-X_{hh}\sin{\varphi}\right).&
\end{aligned}
\label{sixDmel}
\end{equation}
This equivalence reveals an important property of the ansatz used: it has exactly the same form in both the heavy-hole and light-hole exciton coordinates. 

\begin{figure}
\includegraphics[scale=1]{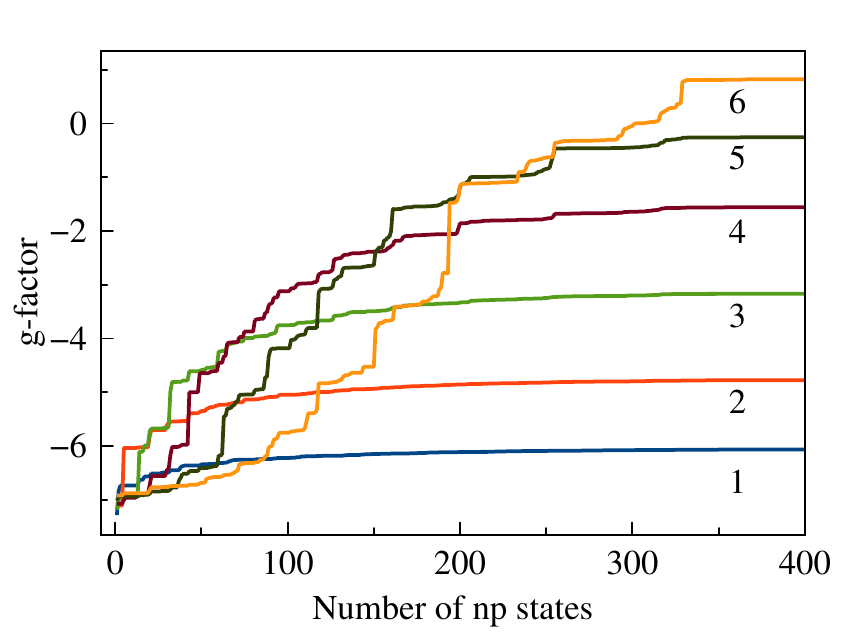}
\caption{The theoretically obtained $g$-factor values versus number of the $p$-like lh-exciton states included in the constrained basis. Number near each curve corresponds to the exciton state number in the 87-nm QW.}
\label{Flo:figure7}
\end{figure}

In section~\ref{modeling}, we also noted that the constrained basis has sufficient number of lh-exciton states. Figure~\ref{Flo:figure7} shows that the calculated $g$-factor values of the five observed hh-exciton states saturate as the number of $p$-like lh-exciton states increases. For the ground state, of about 100 states are sufficient to saturate while, for the fifth state, more than 300 states are needed. According to these saturation data, we have used 400 $p$-like states in the constrained basis.


%
\end{document}